\documentclass[11pt,twoside,onecolumn]{article}
\usepackage[]{amsmath,amssymb}
\usepackage[]{epsfig}
\newcommand{\be}{\begin{eqnarray}}
\newcommand{\ee}{\end{eqnarray}}

\renewcommand{\d}{\mbox{{\rm d}}}
\title{\bf Thermodynamics of a collapsing shell in an expanding Universe}
\author{G.L.~Alberghi\thanks{e-mail: alberghi@bo.infn.it},$\ $
R.~Casadio\thanks{e-mail: casadio@bo.infn.it}
$\ $ and G.Venturi\thanks{e-mail: armitage@bo.infn.it}\\
 \\
{\em Dipartimento di Fisica, Universit\`a di Bologna, and}
\\
{\em Istituto Nazionale di Fisica Nucleare,
Sezione di Bologna, Italy}}
\begin{document}
%
%
\maketitle
\begin{abstract}
We describe the quasi-static collapse of a radiating, spherical shell of matter in
de~Sitter space-time using a thermodynamical formalism.
It is found that the specific heat at constant area and other thermodynamical
quantities exhibit singularities related to phase transitions during the collapse.
\end{abstract}
%
\pagestyle{plain}
\raggedbottom
\setcounter{page}{1}
%
%
%
%
Because of the paradigm of inflation, de~Sitter space-time has recently
become the subject of great research interest.
One aspect of research has been the conjectured dS-CFT
correspondence~\cite{Strominger} which is the analogue of Maldacena's
AdS-CFT conjecture~\cite{AdSCFT} for the case of a space-time with a positive
cosmological constant.
As in the latter case, the ``gravitational'' (dS) side of the correspondence is expected
to yield the macroscopic description of the system, whereas the CFT 
(conformal field theory) should be able to describe the microphysics.
A well known example for the AdS-CFT is given by the microscopical description
of the thermodynamical laws of black holes~\cite{bhs}.
Furthermore, applications of the de~Sitter space Bousso bound~\cite{Bousso}, 
related to the Bekenstein bound for flat space-times, have also drawn much interest.
\par
The dS-CFT correspondence has not yet reached the degree of comprehension of the
AdS-CFT, since the corresponding CFT has not been completely identified.
Therefore, one usually proceeds by studying simplified ``gravitational'' models in order
to obtain general features that the relevant CFT is required to possess.
One such model is given by a thin spherical shell of matter in a de~Sitter
space-time, since it may represent a collection of D-branes forming the so called
D-Sitter space-time of Ref.~\cite{Fabinger}, or can be used as a tool for the study
of the entropy bound in a very simplified physical environment~\cite{Chamblin}.
Finally, one may also consider the ``operational'' approach to black hole entropy,
as illustrated in Ref.~\cite{Pretorius}, which leads one to the conclusion that the
entropy of a black hole is the same as that stored in a shell of matter sitting just
outside its Schwartzschild radius after it has collapsed from infinity.
According to the AdS-CFT correspondence, a black hole is represented by a
thermal state and its formation by the approach to such a state.
A thermodynamical description of the gravitational collapse should thus be
useful in the de~Sitter case as well.
\par
It is in this framework that the study of the collapse of a spherical shell of matter
in a de~Sitter space-time, in analogy with our previous treatment of the
anti-de~Sitter case~\cite{AC}, may be useful and interesting since the former
introduces an additional horizon.
The application of a thermodynamical formalism to describe the collapse
may give important insights on the process and the inclusion of the radiation
coming out of the shell should reveal its importance in the identification of the 
diverse thermodynamical quantities. 
We shall compare our results and extend the evaluation of the specific heats
with the approach of Ref.~\cite{Pretorius} and comment on other
references~\cite{Chamblin,Pretorius}, where different choices for the definition
of the thermodynamical observables are employed.
\par
In particular, we shall examine the collapse of a radiating spherical shell of matter
in de~Sitter space-time with the assumption that collapse is a quasi-static process,
that is, the shell contraction is sufficiently slow so that the system can be described
as evolving through a sequence of equilibrium states.
In fact a radiative process related to the velocity of the collapse (as one would expect)
naturally leads to a quasi-static collapse~\cite{acvv}.
This assumption  has allowed us to introduce a thermodynamical formalism
(see Refs.~\cite{AC,acvv,Alberghi}) to describe the process.
The properties of the system depend on the equation of state, that is
a relation between the thermodynamically independent quantities.
In order to obtain some explicit results we shall consider the general case of
a power-law dependence of the shell temperature (introduced as usual
through the second law of thermodynamics) on the inner horizon radius and also
examine the particular choice corresponding to the Hawking temperature of
the incipient black hole or other choices~\cite{Pretorius}.
We use units for which $\hbar=c=k_{\rm B}=1$, with $k_{\rm B}$ the
Boltzmann constant.
\par
The spherically symmetric space-time we consider is divided into an inner
region and an outer one by a thin massive spherical shell.
The inner region can be expressed in static coordinates as
\be
\d s^2_{\rm i}=-f_{\rm i}(r)\,\d t^2+{\d r^2\over f_{\rm  i}(r)}
+r^2\,\d\Omega^2
\ ,
\ee
and will be taken to be described by a Schwarzschild metric,
so that  $f_{\rm i}(r)=1- 2 \, m / r $ where $m$ is a constant ADM mass.
The outer region, because of the radiation emitted by the shell,
is described by a Vaidya-dS space-time
\be
\d s^2_{\rm o}=-{1\over f_{\rm o}(r,t)}\,\left[
\left({\partial_t M(r,t)\over\partial _r M(r,t)}\right)^2\,\d t^2
-\d r^2\right]+r^2\,\d\Omega^2
\ ,
\ee
with $ f_{ \rm o}(r,t)=1- 2\,M(r,t)/  r - r^2 / \ell^2 $, where $M(r,t)$ is
the Bondi mass and its dependence on the time $t$ is related to the amount
of radiation (energy) flowing out of the shell,
$\partial_t M$ and $\partial_r M$ are the partial derivatives of
$M(r,t)$ with respect to $t$ and $r$ respectively,
and $\ell$ is the dS (cosmological) radius.
\par
Israel's junction equations \cite{Israel} for a static thin
shell located at radius $r=R$, allow us to relate the proper mass of the
shell $E$ to the inner and outer metrics through the equation
\be
E(R,M)=4\,\pi\,R^2\,\rho=
R\,\left(\sqrt{f_{\rm i}(R)}-\sqrt{f_{\rm o}(R)}\right)
\ ,
\label{energy}
\ee
where $\rho$ is the surface energy density,
and $ M $ and $ R $ are the dynamical independent variables
($m$ and $ \ell$ are taken to be fixed).
One may evaluate the surface tension, denoted by $P$, as
\be
P(R,M)&\equiv&{\partial E\over \partial A}
\nonumber
\\
&=&
{1\over 8\,\pi\,R}\,\left[
\sqrt{f_{\rm i}(R)}-\sqrt{f_{\rm o}(R)}
+{1\over\sqrt{f_{\rm i}(R)}}\,{m\over R}
-{1\over\sqrt{f_{\rm o}(R)}}\,\left({M\over R}-{R^2\over \ell^2}\right)
\right]
\ ,
\label{pressure}
\ee
where $A=4\,\pi\,R^2$ is the shell area.
In our thermodynamical description the surface tension and the area 
correspond respectively to the intensive and extensive variables.
The proper mass, on the other hand, corresponds to the internal energy
(see Refs.~\cite{AC,Alberghi}) and Eq.~(\ref{energy}) is a statement of
the first law of thermodynamics ($\d E $ is an exact differential).
\par
Given our previous definitions of the thermodynamical quantities,
one may define the infinitesimal heat flow $\delta Q$
by
\be
\delta Q=\d E-P\,\d A
\ .
\label{I}
\ee
and using the explicit expressions for the pressure and the internal energy
one finds
\be
\delta Q={\d M\over\sqrt{f_{\rm o}(R)}}
\ .
\ee
The above is in agreement with the constraint associated with the continuity
equation for matter in the form
\be
{\d L\over \d\tau}={1\over\sqrt{f_{\rm o}(R)}}\,{\d M\over \d\tau}
\ ,
\label{continuity}
\ee
where  $L$ is the shell luminosity and $\tau$ is the proper time
of an observer sitting on the shell thus confirming our definitions of internal energy
and surface tension ($P$).
\par
It is now possible to introduce a temperature $T$ through the second
law of thermodynamics, that is the existence of the entropy as the exact
differential
\be
\d S={\delta Q\over T}
\ .
\label{entropydiff}
\ee
The temperature appears as an integrating factor (see also Ref.~\cite{Padma1}) 
which must satisfy the integrability condition
\be
{\partial\over\partial R}\,
\left(T\,\sqrt{f_{\rm o}(R)}\right)^{-1}
=0
\ ,
\ee
whose general solution, which exhibits the usual Tolman radial dependence, is
\be
T={B_h(R_h)\over\sqrt{f_{\rm o}(R)}}
\ ,
\label{temperature}
\ee
where $B_h=B_h(R_h)$ is an arbitrary function of the inner horizon radius
$R_h = \sqrt{4/3}\,\ell\,\cos (\theta / 3)$ (where $\cos \theta = -3\,\sqrt{3}\,M\,\ell  $ 
and the outer horizon is given by $R_\ell=\sqrt{4/3}\,\ell\,\cos( \theta /3 + 4\,\pi /3 ) $,
with $ \pi < \theta \le 3\,\pi /2  $ and $ 0 \le 27\,M^2\,\ell^2 < 1$),
which can be used as an independent thermodynamical variable
instead of the Bondi mass $M = (R_h / 2)\, (1-R_h^2 / \ell^2)$.
We finally obtain
\be
\d S=\left(1-3\,{R_h^2\over\ell^2}\right)\,{\d R_h\over 2\,B_h}
\ .
\label{dS}
\ee
\par
Given the expression for the temperature one may evaluate the specific heat
at constant radius
\be
C_R\equiv
T\,\left({\partial S\over \partial T}\right)_R
=
T\,\left({\partial S\over \partial R_h}\right)_R\,
\left({\partial T\over\partial R_h}\right)_R^{-1}
\ .
\label{CR}
\ee
With our definition (\ref{temperature}) for the temperature we have
\be
 C_R = \left[{2\,\ell^2\,B_h'\over \ell^2-3\,R_h^2}
+{B_h\over R\,f_{\rm o}(R)}\right]^{-1}
\ ,
\ee
where $B_h'=dB_h/dR_h$.
Thus $C_R$  shows a singularity for $R$ satisfying
\be
\left(R_h-R\right)\,\left(R^2+R\,R_h+R_h^2-\ell^2\right)
={3\,R_h^2 - \ell^2 \over\left(\ln B_h^2\right)'}
\ .
\label{singularity1}
\ee
\par
The specific heat at constant tension takes the form
\be
C_P&\equiv&T\left({\partial S\over \partial T}\right)_P
\nonumber
\\
&=& 
{T\over 2\,B_h}\,\left(1-3\,{R_h^2\over\ell^2}\right)\,
\left[\left({\partial T\over \partial R_h}\right)_R
-\left({\partial T\over \partial P}\right)_{R_h}\,
\left({\partial P\over \partial R_h}\right)_R\right]^{-1}
\nonumber 
\\
&=&
{ H\, R\,f_{ \rm o}^{7/2}\over
B_h\,
\left({M\over R} - { R^2 \over \ell^2}\right) ^2
-f_{\rm o}^{3/2}\,  
\left(R\,f_{\rm o}\,B_h'\,{\d R_h\over \d M}
+{B_h}\right)\, H} 
\ ,
\label{CP}
\ee
where the function
\be
H (R)={1\over f_{o}^{3/2} }\,
\left(1-{3\,M \over R} + {3\,M^2 \over R^2} - {3\,M\,R \over \ell ^2 }\right)
-{1\over  f_{\rm i}^{3/2} }\,\left(1 - {3\,m \over R} + {3\,m^2 \over R^2} \right)
\ .
\ee
Other thermodynamical quantities of interest, related to the second derivative
of the Gibbs potential~\cite{Zemanski}, are the change of the area with respect to
the temperature for fixed tension $(\partial A / \partial T)_P$ and with respect to
the tension for a fixed temperature $(\partial A / \partial P)_T$.
All the above three quantities show a singular behavior if the denominator of
Eq.~(\ref{CP}) vanishes, that is if there is an $R$ satisfying
\be
&&
{3\,R_h\over 2\,R}\,\left(1-{R_h^2\over\ell^2}\right)\,
\left[1-{R_h\over 2\,R}\,\left(1-{R_h^2\over\ell^2}\right)
+{R^2\over\ell^2}\right]
\nonumber
\\
&&+\left[{f_{\rm o}(R)\over f_{\rm i}(R)}\right]^{3/2}
\left(1-{3\,m\over R}+{3\,m^2\over R^2}\right)
\nonumber
\\
&&=1-{R_h^2\over 4\,R^2}\,
\left[1-{R_h^2\over\ell^2}-{2\,R^3 \over R_h\,\ell^2}\right]^2
\left[1+f_{\rm o}(R)\,
{\ell^2\,\left(\ln B_h^2\right)'\over \ell^2-3\,R_h^2}\right]^{-1}
\ .
\label{singularity2}
\ee
\par
\begin{figure}[t]
\centerline{
\raisebox{2.4cm}{$R_h$}
\epsfxsize=200pt\epsfbox{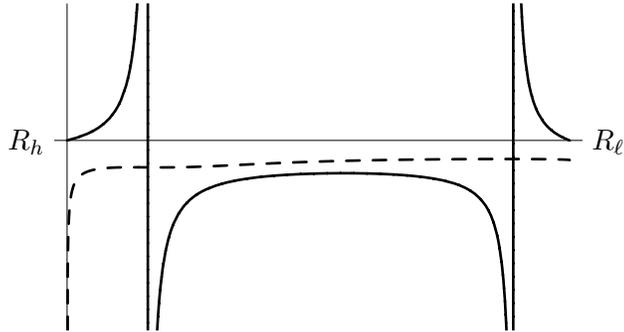}
\raisebox{2.4cm}{$R_\ell$}}
\caption{Behavior of the specific heat at constant
area $C_R$ (solid line) and at constant tension $C_P $ (dashed line)
for a temperature given by Eqs.~(\ref{pow}) and  (\ref{hawkingT}),
and $R$ lying  between the inner ($R_h$) and outer ($R_\ell$) horizons.}
\label{figure1}
\end{figure}

In order to have an explicit expression for the specific heats and to
proceed further in our investigation, we need an equation of state, that
is an expression for the function $B_h$.
Let us examine a rather general case assuming a power-law dependence of
the function $B_h$ on the horizon radius, leading to a temperature
\be
T={1\over \sqrt{f_{\rm o}(R)}}\,{1\over 4\,\pi\,R_h^a}
\ ,
\label{pow}
\ee
with $a$ a constant.
We can now determine the specific heat at constant area
\be
C_R=
-4\,\pi\,f_{\rm o}\,R_h^{a+1}\,
\left(1-3\,{R_h^2\over\ell^2}\right)\,
\left(2\,a\,f_{\rm o}+R_h\,{\partial f_{ \rm o}\over\partial R_h}
\right)^{-1}
\ .
\ee
This implies that $C_R$ diverges for
\be
0&=&
2\,a\,f_{\rm o}+R_h\,{\partial f_{\rm o}\over\partial R_h}
\nonumber
\\ 
&=&
2\,a\,\left(1-{R^2\over \ell^2}\right)
-{R_h\over R}\,\left[(1+2\,a)-(3+2\,a)\,{R_h^2\over \ell^2}\right]
\ .
\label{CR0}
\ee
We have examined numerically the behavior of $C_R$:
for the power-law case it shows two singularities between
the black hole and the cosmological horizons and vanishes on both horizons
(as shown in Fig.~\ref{figure1}).
\par
If we now examine the case for which the temperature is that of a black hole
with horizon radius $R_h$~\cite{hawking}, one has
\be
B_h={1\over 4\,\pi\,R_h}\,\left(1-3\,{R_h^{2}\over \ell^{2}}\right)
\ ,
\label{hawkingT}
\ee
this seems to be the most natural choice if one assumes that at the
end of the collapse the system behaves as if a black hole were being
formed (for an analysis supporting the naturalness of this choice see
Refs.~\cite{Alberghi,Oppenheim,Belgiorno}).
On substituting for $B_h$ in Eq.~(\ref{singularity1}) one
obtains~\footnote{This expression is analogous to Eq.~(A.2) of Ref.~\cite{AC}
where a factor $(1 - 3 R_h^2 / \ell^2 )^{-1}$ was inadvertently omitted.} 
\be
R-{R^{3}\over \ell^{2}}=
{ R_h \over 2\,\left( 1 + {3\,R_h^2 \over \ell^2} \right)}\,
\left(3 - {2\,R_h^2 \over \ell^2} + {3\,R_h^4 \over \ell^2}
\right)
\ ,
\ee
which determines the singularity of the specific heat at constant area.
Its behavior is analogous to that exhibited in the ``power-law'' case of
Eq.~(\ref{pow}) and is again shown in Fig.~\ref{figure1}.
We finally note that for the choice of a Hawking temperature,
the entropy, using Eqs.~(\ref{dS}) and (\ref{hawkingT}),
is given by
\be
S=\int {\delta Q\over T}=\pi\,R_h^2
={1\over 4}\,({\rm horizon\ area})
\ .
\label{SA}
\ee
This expression will exhibit a simple additive property, in the
sense that the entropy of two non-interacting (well separated)
shells will just be the sum of the two entropies, as expected
for usual thermodynamical systems~\cite{Belgiorno}.
Such an additive property also rules out any integration constant
in Eq.~(\ref{SA}).
Concerning the specific heat for constant tension $ C_P $ for a power-law
equation of state~(\ref{pow}) the results are again plotted in
Fig.~\ref{figure1}
and one finds only one singularity on the inner horizon.
The case for a Hawking temperature has a similar behaviour.
\par
\begin{figure}[ht]
\centerline{
\raisebox{1.8cm}{$R_h$}
\epsfxsize=200pt\epsfbox{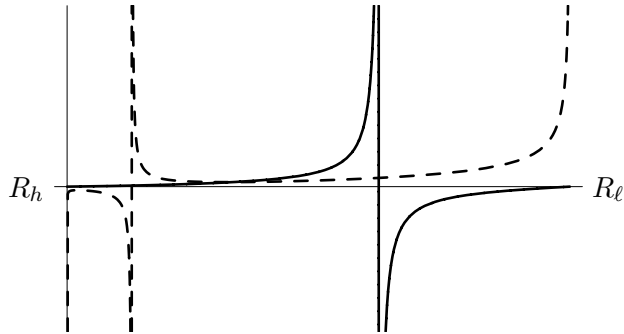}
\raisebox{1.8cm}{$R_\ell$}}
\caption{Specific heat at constant
area $C_R$ (solid line) and at constant pressure $ C_P $ (dashed line) for a 
temperature given by Eq.~(\ref{acctemp}), and $R$ lying 
between the inner and outer horizons.}
\label{figure2}
\end{figure}

Let us briefly compare our results with those of other approaches.
In  Ref.~\cite{Pretorius} a basic statement is that the temperature attributed to the
shell is that of a static observer standing just outside the shell, that is
\be
T (R) ={a(R) \over 2\,\pi }
=
{1 \over  {4\,\pi\,\sqrt{f_{\rm o} (R)} } }\,
{\partial f  _{\rm o}(R) \over\partial R}
\ ,
\label{acctemp}
\ee 
where $a(R)$ is the acceleration of a static observer.
A chemical potential must be introduced in order to obtain the entropy
as an exact differential.
Indeed, by defining the temperature as related to the acceleration of a
static observer on the shell, one is led to the definition  of the entropy
differential
\be
\d S = \beta\,dM + \beta\,P\,\d A - \alpha\,\d N
\ ,
\ee
where $ \beta = T ^{-1}$, $\alpha = \mu / T$ with $\mu $ the chemical
potential, and $N$ can be interpreted as the number of particles in the shell.
Using the Gibbs-Duhem relation~\cite{Pretorius}, one has
\be
S = \beta\, (M + P\,A) - \alpha\,N
\ ,
\ee
or 
\be
n\,\d\alpha = \beta\,\d P + (\rho + P)\,\d \beta
\ ,
\ee
with $ n=N/A$.
This definition, after some calculations, gives an explicit expression for the 
entropy density as
\be
 s = S/ A = \beta\, P = {1 \over 4}\,\left( 1 - {1 \over \gamma ^2 } \right)
\ ,
\label{israelentropy}
\ee
with 
\be
\gamma^2 = {1 \over 8\,\pi\,\rho\,\sqrt{f_{o} (R) } }\,
\left( {\partial f_{\rm o} \over \partial R } \right)   
\ ,
\ee
and we note that $S$ becomes $1 / 4$ of the horizon area as the shell reaches
the inner horizon radius.
Further we observe that, on choosing the temperature as suggested in
Ref.~\cite{Pretorius} and using the definition~(\ref{CR}), one can see that
there is only one singularity in $C_R$ (see Fig.~\ref{figure2}).
\par
Finally we mention the approach of Ref.~\cite{Chamblin} where
an expression for the entropy is derived as a integral over a
``Bousso lightsheet''~\cite{Bousso} within the framework of the approach
to the entropy bounds for asymptotically de~Sitter space-times.
In this formalism one can obtain the expression for the entropy as a function
of the shell radius.
The result obtained for the entropy coincides with our definition of the internal
energy $ E(R,M) $.
This leads to an expression for the temperature as given by
Eq.~(\ref{entropydiff}),
\be
T= { \sqrt{f_{\rm o} (R)} \over \pi\,R }
\ ,
\ee
which does not seem to be related to any property of the shell or of the space-time
and does not lead to any significant thermodynamical description of the collapse
process.
Indeed we note that this temperature vanishes as the shell approaches the black
hole horizon.
From the above considerations the only approach compatible with ours 
is that of Ref.~\cite{Pretorius} where, however, one does not have a radiative
metric and the temperature is determined by the Unruh effect.
\par
Let us then illustrate our approach before comparing our results with those
of Ref.~\cite{Pretorius}.  
We have analyzed the thermodynamical behavior for the
collapse of a radiating shell in de~Sitter space-time, under the assumption
that the evolution consists of a succession of equilibrium states, that is the
process is quasi-static.
We note that a radiative process related to the velocity of the collapse
(as one would expect) naturally leads to a  quasi-static collapse~\cite{acvv}.
On identifying the internal energy (\ref{energy}) and surface tension
(\ref{pressure}) of the shell, we were able to evaluate the specific heats at
constant area ($C_R$) and tension ($C_P$)
and other related thermodynamical quantities when the temperature is
given by a power-law of the horizon radius as in Eq.~(\ref{pow}) or by
the Hawking value.
Since we are just considering the ``gravitational'' part of the dS-CFT
correspondence, we can only study the macroscopic structure of the
system and obtain general results.
Indeed from Fig.~\ref{figure1} we see that $C_R$ vanishes on the
black hole and cosmological horizons and exhibits two singularities
located between them.
This structure is clearly related to the existence of two horizons,
as can be inferred by comparing with the anti-de~Sitter case of Ref.~\cite{AC}. 
In the case of Ref.~\cite{Pretorius} there is only one singularity between
the horizons and $C_R$ again vanishes on the horizons.
The sign of $C_R$ is generally associated with thermodynamical stability,
in particular a negative sign implies instability
(see Ref.~\cite{Vakkuri}).
One can also argue that the singularities of $C_R$ correspond to second
order phase transitions insofar as the collapse (and state of the shell)
is expected to be continuous~\cite{Zemanski}.
Naturally, should the internal dynamics of the shell exhibit a discontinuity,
the above considerations are not valid and one could also have a  first order
phase transition.
However, as we mentioned previously, such considerations would require
the knowledge of the shell microscopic structure which should be
described by a CFT.
\par
Concerning our results for $C_P$, we note that often phase transitions are
associated with the behaviour of the Gibbs potential since they occur for constant
temperature and pressure (and chemical potential)~\cite{Zemanski}.
Indeed we found that all the second derivatives of the Gibbs potential,
and in particular $C_P$, exhibit a singularity at the same point
(see Eq.~(\ref{singularity2})) implying a second order phase transition at the
inner horizon (this means that there is no change of entropy as the shell
approaches the inner horizon).
Let us at this point note the difference between our case and that of
Ref.~\cite{Pretorius}.
Even with a definition of the temperature as being related to the acceleration 
of a static observer, the specific heat at constant area shows a singularity
(see Fig.~\ref{figure2}).
However, the nature of the transition is not clear in this case, because of the
introduction of a chemical potential.
\par
Let us end with some comments about the Bekenstein-Bousso bound,
which, for our system, states that 
\be
S \le 2\, \pi\, E\, R
\ ,
\label{bb}
\ee
where $E$ is the total energy of the system and $R$ is its maximum dimension.
Eq.~(\ref{bb}) is easily satisfied both for Eq.~(\ref{SA}) and (\ref{israelentropy}),
whereas for the power-law case of Eq.~(\ref{pow}) one can use the
bound (\ref{bb}) in order to constrain the integration constant for the entropy.
%
%
%
%

%
%
\end{document}